\documentclass[a4paper,10pt]{article}
\usepackage[utf8]{inputenc}
\usepackage{graphicx}
\usepackage{epsfig}
\usepackage{amsfonts}
\usepackage{amsmath}
\usepackage{amssymb}
\usepackage{subfigure}
\usepackage{authblk}
\usepackage{color}
\usepackage{multirow}
\usepackage{array}
\usepackage{bm}
\usepackage{hyperref}

\usepackage{mathtools} 

\def\correspondingauthor{\footnote{Corresponding author.}}

\title{Stochastic Learning in Kolkata Paise Restaurant Problem: Classical \& Quantum Strategies}
\author[1,2,3]{Bikas K Chakrabarti \thanks{bikask.chakrabarti@saha.ac.in}}
\author[4]{Atanu Rajak \thanks{atanu.physics@presiuniv.ac.in}}
\author[5]{Antika Sinha\correspondingauthor{} \thanks{antikasinha@gmail.com}}
\affil[1]{Condensed Matter Physics Division, Saha Institute of Nuclear Physics, 1/AF Bidhannagar, Kolkata 700064, India}
\affil[2]{S. N. Bose National Centre for Basic Sciences, Kolkata 700106, India}
\affil[3]{Economic Research Unit, Indian Statistical Institute, Kolkata 700108, India}
\affil[4]{Department of Physics, Presidency University, Kolkata 700073, India}
\affil[5]{Department of Computer Science, Asutosh College, Kolkata 700026, India}

\begin{document}

\maketitle

\begin{abstract}
We will review the results for
stochastic learning strategies,
both classical (one-shot and iterative) and quantum (one-shot only), for optimizing
the available many-choice resources among a
large number of competing agents, developed
over the last decade in the context of the
Kolkata Paise Restaurant Problem. Apart from a
few rigorous and approximate analytical results,
both for classical and quantum strategies,
most of the interesting results on the phase
transition behavior (obtained so far for the
classical model) using classical Monte Carlo
simulations. All these, including the 
applications to computer science (job or 
resource allotments in Internet-of-Things), transport engineering 
(on-line vehicle hire problems), operation research (optimizing efforts for delegated search problem,  efficient solution of Travelling Salesman problem), etc will be discussed.
\end{abstract}

\section{Introduction}

Game theory was initially evolved
to investigate different strategic
situations with competing players~\cite{morgenstern1953theory}.  Of late, the concept of game
theory is being applied to
different statistical  events to
measure the success rate when one’s
success depends on the choice of the
other agents. The game of Prisoners’
dilemma (see e.g.,~\cite{kuhn2019prisoner}) is a popular
example where two non-communicating
(or non-interacting) agents choose
their actions from two possible
choices. It is a two-person, two-choice,
one-shot (one time decision) game. The
Nash equilibrium (see e.g.,~\cite{osborne1994course})
solution employs the strategy, where
the other player can not gain from any
of the choices, and both the players
necessarily defect. However,  this is
not a Pareto optimal solution (see
e.g.,~\cite{lockwood2008pareto}), where no change in the
decision can lead to a gain for one
player without any loss of the other.
This problem has been used to model
many real life problems like auction
bidding, arms races, oligopoly pricing,
political bargaining, salesman effort
etc.

The minority game theory (see e.g.,~\cite{challet2004minority}) generalizes this idea of a very
large number of non-communicating
players with with two choices for
each of them. As the name suggests,
the players who make the minority
group choice (at any time) receive
a payoff. This game is not an
one-shot game and the players learn
from their previous mistakes (loss of
payoffs) and continuously try upgrade
their respective respective strategies
to gain the payoffs and they (the
society as a whole) learn  collectively
to reach a level of maximum efficiency,
where no one  can improve their payoff
any further. A phase  transition
(see e.g.,~\cite{challet2004minority}) occurs at a critical
value the memory size (number of
distinct strategies individually
remembered; assumed to be the same
for all the players~\cite{stanley1972introduction}) and the number
of players and the socially optimal
learning tome diverges at this critical
point. The game has many important
application of social dilemmas,
including a decision of  making
investment in  a stock market;
over-crowding of the agents any
day due to decision either  buying
or selling a particular stock in a
financial market can lead to loss
for the majority players!

The minority game is further
generalized for many choices in
addition to many players (as the
minority game) in the Kolkata Paise
Restaurant (KPR) game theory,
introduced by Chakrabarti~\cite{chakrabarti2007kolkata} and
Chakrabarti et al.~\cite{chakrabarti2009kolkata} (for a recent review see e.g.,
 Chakrabarti et al.~\cite{chakrabarti2017econophysics}). The KPR game is also  an
iterative game, played by the agents
or players without any interaction or
communication between each other.

In Kolkata, long back, there  were
very  cheap  and fixed price ``Paise
Restaurants''  (also  called ``Paise
Hotels'';  Paise was the  smallest Indian  coin) which
were very  popular  among  the daily
laborers  in  the  city. During lunch
hours, these laborers used to walk
down (to  save  the  transport costs)
from their place of work to one of
these restaurants. These  Paise
Restaurants  would  prepare  every
day a fixed (small)  number of such
dishes, and if several groups  of
laborers  would arrive  any day to
the same restaurant, only one group
perhaps  would  get  their lunch
and  the  rest  would miss  lunch
that  day. There were no cheap
communication means (mobile phones)
for mutual interactions,  in order to 
decide about the  respective
restaurants of the day. Walking  down to the
next restaurant would mean failing
to report  back  to  work  on time!
To  complicate  this collective
learning and decision  making problem,
there  were indeed  some well-known
rankings of these restaurants, as
some of them would  offer  tastier
items  compared  to  the  others (at
thesame  cost,  paisa,  of  course)
and  people  would  prefer  to choose
the  higher  rank  of  the  restaurant,
if  not  crowded! This ``mismatch'' of
the choice and the consequent decision
not only creates inconvenience for the
prospective customer (going  without
lunch),  would  also  mean  ``social
wastage'' (excess unconsumed food,
services or supplies somewhere).

A similar problem arises when the
public administration  plans  and
provides  hospitals  (beds)  in
different  localities,  but  the
local  patients  prefer  ``better''
perceived  hospitals  elsewhere.
These  ``outsider''  patients then
compete  with  the  local  patients
and  have  to  choose other  suitable
hospitals elsewhere. Unavailability
of the  hospital  beds  in  the
over-crowded hospitals may be
considered as insufficient service
provided by the administration, and
consequently the unattended potential
services  will  be  considered as
social  wastage. This  kind  of  games,
anticipating  the  possible strategies
of the other players and acting
accordingly, is very common  in
society.  Here,  the number of choices
need not  be  very  limited  (as  in
the  standard  binary-choice formulations
of most  of  the  games)  and  the number
of players  can  be  truly  large! Also,
these  are  not  necessarily one  shot
games,  rather  the  players  can  learn
from  past mistakes  and  improve  on
the  selection  of  their  strategies
for  the  next  move.  These  features
make  the  games extremely intriguing and
also versatile, with major collective
or  social  emerging  structures.

The KPR  problem  seems to  have  a
trivial  solution:  suppose  that
somebody,  say  a dictator (who is not
a player), assigns a restaurant to
each person  and  asks  them  to  shift
to  the  next  restaurant cyclically,
on  successive  evenings.  The  fairest
and  most efficient solution: each
customer gets food on each evening
(if  the  number  of  plates  or
choices  is  the  same  as  that  of
the  customers  or  players)  with
the  same  share  of  the rankings
as  others,  and  that  too  from  the
first  evening (minimum  evolution time).
This,  however,  is  not  a  true
solution of the KPR problem, where each
customer or agent decides  on  his  or
her  own  every  evening,  based  on
complete  information  about  past events.
Several recent applications of the
classical KPR strategies to Vehicle for
Online Hire problem~\cite{martin2017extending,martin2017vehicle}, resource
allocation problem in the context of
Internet-of-Things~\cite{park2017kolkata}, developing
a different strategy for solving the the
Traveling Salesman Problem~\cite{kastampolidou2021dkprg}, etc
have been made.

For the last few decades, quantum game
theory  has been developed, promising
more success over classical strategies
~\cite{meyer1999quantum,eisert1999quantum,benjamin2001multiplayer,marinatto2000quantum,piotrowski2003invitation,bleiler2008formalism,salimi2009investigation,landsburg2011quantum}.
This is an interdisciplinary approach that
connects three different fields: quantum
mechanics, information theory and game
theory in a concrete way. Quantum game
theory offers different protocols that
are based on the uses of quantum
mechanical  phenomena like entanglement,
quantum  superposition, and interference
arising due to wave mechanical aspects
of such systems. In context of game
theory, quantum strategies are first
introduced in two papers  by Meyer~\cite{meyer1999quantum}
and by Eisert et al.~\cite{eisert1999quantum}
 where they showed that a player
performing a quantum move wins against
a player performing a classical move
regardless of their classical
choices. The advantage of a quantum
strategy over classical one has been
specifically investigated in~\cite{eisert1999quantum} 
for the case of Prisoners’ dilemma.
This idea is generalized for multiple
players by Benjamin and Hayden~\cite{benjamin2001multiplayer} 
with a specific solution  for four
players. The authors here introduced
quantum minority game where they
showed that an entanglement shared
between the players promises better
performance of quantum strategy over
the classical one. Chen et al.~\cite{chen2004n} 
further extends this result of quantum
minority game for $N$-players.
     
Since then, different aspects of multi-player
minority games are being studied extensively. As
already mentioned, the KPR problem is a
minority game with a large number of choices
for each of the players, who are also equally
large in number. In 2011, Sharif and Heydari~\cite{sharif2011quantum} 
introduced the quantum version of the
KPR game, with a solution for three agents
and three choices. This study was later
extended by Sharif and Haydari~\cite{sharif2012strategies} in 2012, Ramzan~\cite{ramzan2013three} and Sharif and Haydari~\cite{sharif2013introduction} in 2013   
 for the three and multi-player quantum
minority games, including the quantum KPR
games (essentially one-shot solutions). For 
a detailed discussion see Chakrabarti et al.~\cite{chakrabarti2017econophysics}.

We review here the statistics of the KPR
problem employing  both classical and quantum
strategies. The article is organized as follows. In Sec.~\ref{classical_strategies}, we describe the classical strategies of KPR game and show that there exists a phase transition when the number of customers is less than the number of restaurants. We also discuss there the possible ways by which we can minimize the social wastage fraction. In Sec.~\ref{quantum_games}, we first discuss about the general setting of quantum games and then provides a flavour of two game theoretical problems, such as, Prisoners' dilemma and minority game in the context of both classical and quantum strategies. In Sec.~\ref{quantum_KPR}, we introduce quantum version of KPR problem. We review here the results of one shoot quantum KPR problem with three players and three choices by Sharif and Haydari~\cite{sharif2011quantum,sharif2012strategies,sharif2012introduction}, and Ramzan~\cite{ramzan2013three}. We show that using quantum strategies one can gain in payoff by $50\%$ compared to the classical strategies for one shoot KPR game with three players and three choices. We also discuss about the effect of entanglement and decoherence (or loss of phase coherence) in finding the expected payoff of a player for the mentioned game 
problem.



\section{ Statistics of KPR game: Classical strategies}
\label{classical_strategies}
Let us consider the KPR game with $N$ restaurants and $\lambda N$ non-communicating players (agents or customers). We assume that every day or evening or time ($t$), each restaurant prepares only one dish (generalization to a larger number would not affect the statistics of the game). As discussed, every time $t$, the objective of each of the player is to choose one among $N$ restaurant such that she will be alone there in order to get the only dish. If some restaurant is visited by more than one customer, then the restaurant selects one of them randomly and serve the dish to her; thus rest of the visitors there would remain unhappy by starving that evening.

Let us consider first the random choice (no learning) case where each player chooses randomly any of the restaurants.
 Then the probability $P$ of choosing one restaurant by $n~(\le N)$ players is

 \begin{equation}\label{eq_KPR_I} 
P(n) = \binom{\lambda N}{n} {p}^{n} {(1-p)}^{\lambda N - n};~p=\frac{1}{N}. 
\end{equation}

\noindent
In the case $N$ goes to infinity, we get

\begin{equation}\label{eq_KPR_II}
P(n) = \frac{\lambda^n }{n!} exp(-\lambda). 
\end{equation}
Hence, the fraction $P(n=0)$ of restaurants not chosen by any customer is $exp(-\lambda)$. The fraction of restaurants chosen by at least one customer on any evening is therefore gives the utilization fraction~\cite{chakrabarti2009kolkata}

\begin{equation} \label{eq_KPR_III}
 f = 1-exp(-\lambda).
\end{equation}

\noindent
If $N$ agents (where $\lambda = 1$) randomly choose and visit any one among $N$ restaurants then utilization fraction $f$ becomes $1-exp(-1)\simeq 0.63$. Since there is no iterative learning for this case, every time the utilization fraction will be about $63\%$ starting from the first day (convergence time $\tau$ = 0).

It may be noted that a dictated solution to the KPR problem is simple and very efficient from the first day. The Dictator is not a player in the game and asks the players to form a queue (with periodic boundary condition), visit a restaurant according to her respective positions in that queue and continue shifting by one step every day. Every player gets a dish and hence the steady state ($t$-independent) social utilization fraction $f$ becomes maximum (unity) from the first day ($\tau = 1$). This dictated solution is applicable even when the restaurants have ranks (agreed by all the customers) i.e. agents have their preferences over the restaurants. Thus the dictated solution is very efficient in achieving maximum utility from the first day $(f = 1, \tau = 1)$. However no choice of the individual is considered here and in a democratic set-up no such a dictatorial strategy is acceptable.

We now consider the case where the players try to learn and update their strategies of choosing a restaurant to avoid overcrowding the chosen restaurant. As already discussed, we measure the social utilization fraction $f(t)$ on any day  $t$ as

\begin{equation}\label{eq_KPR_IV}
f(t) = \sum_{i=1}^{N} [\delta(n_i(t))/\lambda N],
\end{equation}

\noindent
where $\delta(n) = 1$ for $n \ge 1$ and $\delta(n) = 0$ for $n=0$; $n_i(t)$ denotes the number of customers arriving at the $i$th (rank if customer choice is considered) restaurant on $t$ th evening. The goal is to learn collectively towards achieving $f(t)$ = 1 preferably in finite convergence time $\tau$, i.e., $f(t)$ = 1 for $t \ge \tau$, where $\tau$ is finite.

Earlier studies (see
e.g.,~\cite{chakrabarti2009kolkata,chakrabarti2017econophysics,ghosh2010kolkata,ghosh2010statistics,ghosh2012phase,sinha2020phase}) had proposed several learning strategies for KPR game. 
In references~\cite{ghosh2010statistics,ghosh2012phase,sinha2020phase,chakrabarti2021development}, the authors had studied several stochastic crowd avoidance learning strategies leading to increased utilization fraction (compared to the random choice case Eq.~($\ref{eq_KPR_III}$)). In some of the cases this is achieved ($f=1$) at a critical point~\cite{stanley1972introduction} where $\tau$ goes to infinity due to critical slowing down. 


Here we discuss numerical (Monte Carlo) results for the statistics of KPR game where $\lambda N$ ($\lambda > 0$) customers choose one among $N$ restaurants following a strategy discussed next. On day $t$, an agent goes back to her last day's visited restaurant $k$ with probability 
\begin{subequations}
\begin{gather}
  p_k(t) = {[n_k(t-1)] }^{-\alpha};~\alpha > 0 \label{eq_KPR_V}\\
 \shortintertext{and chooses a different restaurant $(k^{'}\neq k)$ among any of the ($N-1$) neighboring restaurants,  with probability}
p_{k^{'}}(t) = (1-p_k(t))/(N-1). \label{eq_KPR_VI}
\end{gather}
\end{subequations}

\noindent
These ``learning'' strategies employed by the players, for the choice of restaurants placed on different dimensional ($d$) lattices. In infinite dimension (mean field case) the restaurant indices $k$ and $k^{'}$ in Eq.~($\ref{eq_KPR_V}$,~$\ref{eq_KPR_VI}$) run from $1$ to $N$ of the lattice. For finite dimensions $k$ runs from $1$ to $N$ while $k^{'}$ corresponds to the nearest neighbour of the $k$ th restaurant on the lattice.

Authors in~\cite{ghosh2012phase} had studied crowd dynamics with $\alpha=1, \lambda=1$ in infinite, $2d$, $1d$  lattice structure of restaurants. KPR dynamics for $\alpha\le1, \lambda=1$ had been studied in~\cite{sinha2020phase} for infinite, $3d,2d,1d$ lattice structure of restaurants. Phase transition behaviours are observed for $\alpha$ near $\alpha_c=0_+$ for infinite, $3d$ and $2d$ lattice structure of the restaurants. 
The steady state statistics are studied when the utilization fraction $f(t)$ remains the same (within a predefined margin) for further iterations. The steady state wastage fraction $(1-f)$ and the convergence time $\tau$ for reaching the steady state are found to vary with $\Delta \alpha \equiv |\alpha-\alpha_c|$ as $(1-f)\sim \Delta \alpha^{\beta}$ and $\tau \sim \Delta \alpha^{-\gamma}$ with $\beta \simeq 0.8,0.87,1.0$  and $\gamma \simeq 1.18,1.11,1.05$ in infinite-dimension, $3d$ and $2d$ lattice structures respectively. Results of $1d$ lattice structure is found to be trivial unlike other dimensions and no phase transition ($f$ reaches unity with no divergence in $\tau$) is seen for any $\alpha > 0$. 

\begin{figure*}[htb]
\begin{center}
\includegraphics[width=12.5cm]{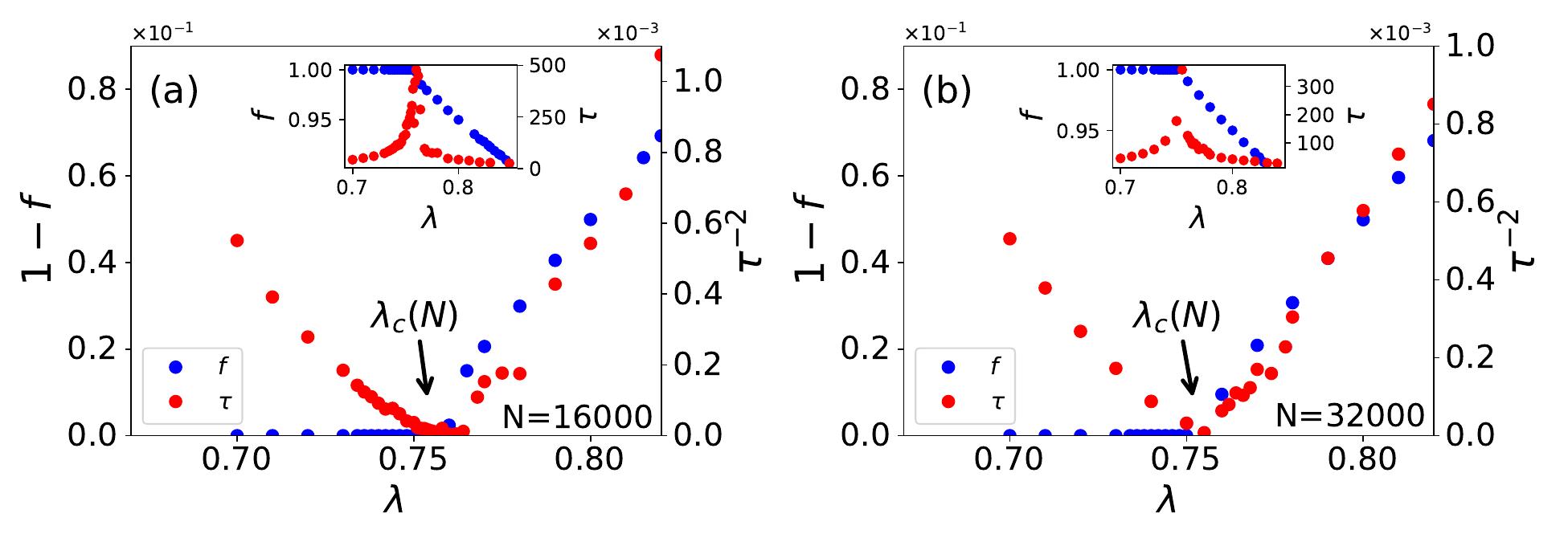} 
\end{center}
\caption{ Plots of steady state convergence behaviour in infinite dimensional lattice; Social wastage fraction $(1-f)$ and convergence time $\tau$  against customer fraction $\lambda$. Observed power laws are $(1-f) \sim \Delta\lambda^{\beta}$ where $\beta = 1.0\pm0.02$ and $\tau \sim \Delta\lambda^{-\gamma}$ where $\gamma = 0.5\pm0.02$. Here $\Delta \lambda \equiv |\lambda-\lambda_c(N)|$. The insets show the variation of $f$ and $\tau$ against $\lambda$, showing the diverging behaviour of $\tau$ where $f$ reaches unity at $\lambda = \lambda_c(N)$. (a) for $N=16000$ and (b) for $N=32000$.
 }
\label{fig_I}
\end{figure*}

\begin{figure*}[htb]
\begin{center}
\includegraphics[width=7cm]{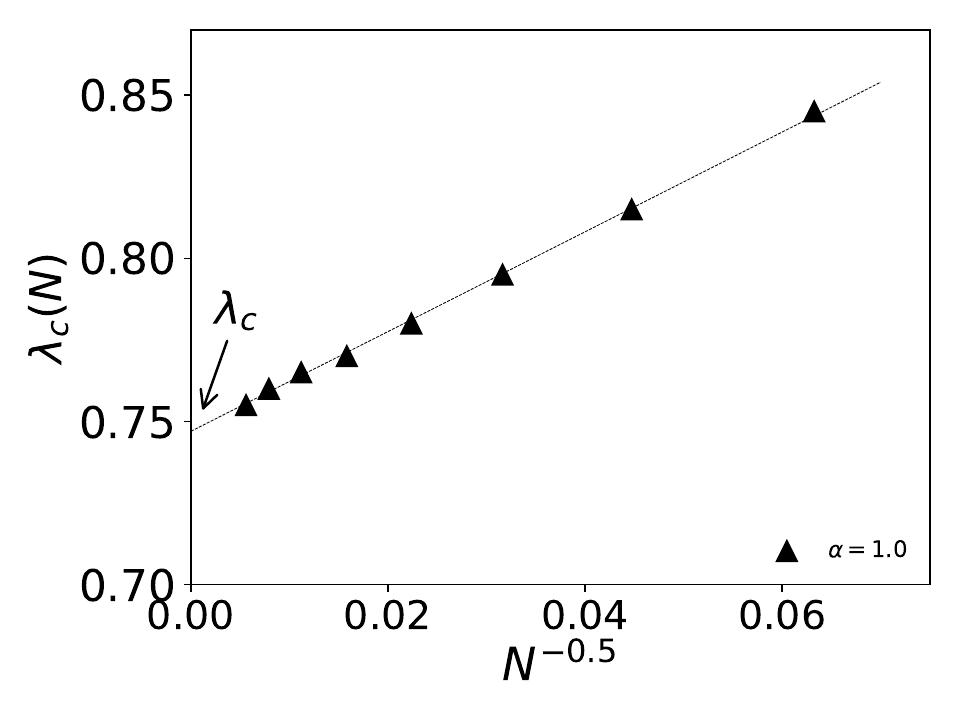} 
\end{center}
\caption{ Extrapolation study of the effective finite size dependent critical density of customers $\lambda_c(N)$. The system size dependence is fitted to  $\frac{1}{\sqrt{N}}$ and we get $\lambda_c=0.74\pm0.01$ for $\lambda_c \equiv \lambda_c(N\to \infty)$. 
 }
\label{fig_II}
\end{figure*}

Here we discuss the numerical results for the Monte Carlo studies on steady state statistics of the KPR game dynamics for general $\alpha$ and $\lambda$ cases. In the case where $\alpha = 1$ and $\lambda$ is $<1$, we find power law fits for social wastage fraction $(1-f) \sim \Delta\lambda^\beta$ and convergence time $\tau \sim \Delta\lambda^{-\gamma}$ with $\beta=1.0\pm0.02$ and $\gamma=0.5\pm0.02$ in infinite dimension with $\Delta \lambda \equiv |\lambda - \lambda_c|$, where $\lambda_c=0.74\pm 0.01$ (see Figs~\ref{fig_I},~\ref{fig_II}). 
For finite size $N$, we observe the effective critical point $\lambda_c(N)$ for which the finite size scaling (see  Fig.~\ref{fig_II}) gives best fit for $d\nu = 2.0$ and $\lambda_c = 0.74\pm 0.01$.

A crude estimate of the mean field value of $\lambda_c$ can be obtained as follows. Here $\lambda N$ agents choose every day among the $N$ restaurants. Hence the probability of any restaurant to be chosen by a player is $\frac{\lambda N}{N}=\lambda$ and the fraction of restaurants not visited by any player will be $(1-\lambda)$. In the steady state, the number of players $n$ choosing any restaurant can be 0, 1, 2, 3,... .
 If we assume the maximum crowd size at any restaurant on any day to be 2, then the probability of those restaurants to go vacant next day will be $(\frac{1}{2})^{2}$. Hence the critical value $\lambda_c$ of $\lambda$ in the steady state will be given by 

\begin{equation}\label{eq_KPR_VII}
 1-\lambda_c=\frac{1}{4},
\end{equation}

\noindent
giving $\lambda_c\simeq 0.75$ . For more details see~\cite{ghosh2012phase}.


\section{Quantum Games}
\label{quantum_games}
In the setting of quantum games, the $N$ different choices of any arbitrary player or agent are encoded in the basis states 
of a $N$-level quantum system that acts as a subsystem with $N$-dimensional Hilbert space. The total system 
for $M$ ($=\lambda N$ as defined in Sec.~\ref{classical_strategies}) players or agents
can be represented by a state vector in a $\prod_{i=1}^M{\rm dim}(\mathcal{H}_{\mathcal{L}_i})$ dimensional Hilbert space 
$\mathcal{H}_{\mathcal{L}}=\mathcal{H}_{\mathcal{L}_M}\otimes \mathcal{H}_{\mathcal{L}_{n-1}}\cdots \otimes \mathcal{H}_{\mathcal{L}_1}$,
where $\mathcal{H}_{\mathcal{L}_i}$ is the Hilbert space of the $i$-th subsystem. 
The different subsystems are distributed among the players and the initial state of the total system is chosen so that 
the subsystems become entangled. The players do not communicate among each other before choosing a strategy. 
A strategy move in quantum games is executed by the application of local operators associated to each player on the quantum state.
The players do not have access any other parts of the system except their own subsystems. In addition, no information 
is shared between the players exploiting the quantum nature of the game. The quantum strategies are indeed 
the generalized form of classical strategies with $s_i \in S_i \; \Rightarrow \;U_i \in \textrm{S}(N_i)$, where 
the set of permitted local quantum operations $\textrm{S}(N_i)$ is some subset of the special unitary group $\textrm{SU}(N_i)$.

We will now describe different steps of the quantum game protocol~\cite{sharif2012introduction}. The game starts with an initial entangled state $|\psi_n\rangle$ shared 
by different players. We have considered the subsystems of the same dimension $N$ that indeed denotes the number of pure strategies 
available to each player. The number of subsystems is equal to the number of players. It can be thought that $|\psi_n\rangle$ 
has been prepared by a referee who distributes the subsystems among the players. By choosing a unitary operator 
$U$ from a subset of SU($N$), the players apply that on their subsystems and, the final state is given by 
\begin{equation}
\rho_{\rm fin}=U\otimes U\otimes \cdots \otimes U \rho_{\rm in} U^{\dagger}\otimes U^{\dagger}\otimes \cdots \otimes U^{\dagger},
\end{equation}
where $\rho_{\rm in}$ and $\rho_{\rm fin}$ are the initial and final density matrix of the system, respectively.
Due to the symmetry of the games and, since the players do not communicate among themselves, they are supposed to do 
the same operation. The advantage of quantum game over classical one is that it reduces the probability of collapsing 
the final state $\rho_{\rm fin}$ to the basis states that have lower or zero payoff $\$$. Since quantum mechanics is a 
fundamentally probabilistic theory, the only notion of payoff after a strategic move is the expected payoff. 
To evaluate the expected payoffs, the first step is to define a payoff operator $P_i$ for an arbitrary player $i$ and
that can be written as
\begin{equation}\label{payoffoperator}
P_i = \sum_j \$^j_i|\alpha^{j}_i\rangle \langle\alpha^{j}_i|,
\end{equation}
where $\$_i^j$ are the associated payoffs to the states $|\alpha^{j}_i\rangle$ for $i$-th player. 
The expected payoff $E_i(\$)$ of player $i$ is then calculated by considering the trace of the product of the final 
state $\rho_{\rm fin}$  and the payoff-operator $P_{i}$,
\begin{equation}
E_{i}(\$)=\mathrm{Tr\left(\mathit{P}_{i}\rho_{\rm fin}\right)}.
\label{expect_payoff}
\end{equation}

The Prisoners' dilemma is a game with two players and both of them have two independent choices.
In this game, two players, Alice and Bob choose to cooperate or defect without sharing any prior information about their 
actions. Depending on their combination of strategies, each player receives a particular payoff. 
Once Bob decides to cooperate, Alice receives payoff $\$_A=3$ if she also decides to cooperate, and she receives $\$_A=5$ 
if she decides to defect. On the other hand, if Bob sets his mind to defect, Alice receives $\$_A=1$ by following Bob and,
$\$_A=0$ by making the other choice. It reflects that whatever Bob decides to choose, Alice will always gain if she decides 
to defect. Since there is no possibility of communications between the players, the same is true for Bob. 
This leads to a dominant strategy when both the players defect and they both have payoff $\$_{\rm A/B}=1$. 
In terms of game theory, this strategy of mutual defection is a Nash equilibrium, because none of the players 
can do better by changing their choices independently. However, it can be noted that this is not an efficient solution. 
Because there exists a Pareto optimal strategy when both the players cooperate, and they both receive $\$_{\rm A/B}=1$. 
This gives rise to a dilemma in this game. After a few decades, the quantum version of this game is introduced by 
J. Eisert, M. Wilkens, and M. Lewenstein in $1999$~\cite{eisert1999quantum}. In the quantum formulation, the possible outcomes of classical 
strategies (cooperate and defect) are represented by two basis vectors of a two-state system, i.e., a qubit. 
For this game, the initial state is considered as a maximally entangled Bell-type state, and the strategic moves for 
both the players are performed by the unitary operators from the subset of $\textrm{SU(2)}$ group. 
In this scenario, a new Nash equilibrium is emerged in addition to the classical one, i.e., when both the players 
choose to defect. For the new case, the expected payoffs for both the players are found to be $E(\$_A)=E(\$_B)=3$. 
This is exactly the Pareto optimal solution for the classical pure strategy case. In the quantum domain, this also 
becomes a Nash equilibrium. Thus considering a particular quantum strategy one can always get a advantage over 
a classical strategy.

The above game is generalized for multiple players with two choices in the minority game theory. In this game, $n$ 
non-communicating agents independently make their actions from two available choices, and the main target of the players 
is to avoid the crowd. The choices are then compared 
and the players who belong to the smaller group are rewarded with payoff $\$=1$. If two choices are evenly distributed,
or all the players make the same choice, no player will get any reward. To get the Nash equilibrium solution, the palyers 
must choose their moves randomly, since the deterministic strategy will lead to an undesired outcome where all the 
players go for the same choice. In this game, the expected payoff $E(\$)$ for a player can be calculated by the ratio 
of the number of outcomes that the player is in the minority group and the number of different possible outcomes. 
For a four-player game, the expected payoff of a player is found to be $1/8$, since each player has two minority 
outcomes out of sixteen possible outcomes. A quantum version of this game for four players was first introduces by 
Benjamin and Hayden in $2000$~\cite{benjamin2001multiplayer}. They have shown that the quantum strategy provides a better performance 
than the classical one for this game. The application of quantum strategy reduces the probability of even distribution 
of the players between two choices, and this fact is indeed responsible for the outerperformance of quantum strategy 
over classical one. The quantum strategy provides an expected payoff $E(\$)=1/4$ for four-player game which is twice 
of the classical payoff.

\section{Quantum KPR problem}
\label{quantum_KPR}
As already mentioned, the Kolkata restaurant problem is a generalization of the minority game, where $\lambda N$ non-communicating agents or players generally 
have $N$ choices. The classical version of the KPR problem has been discussed in Sec.~\ref{classical_strategies}. This problem is also 
studied in quantum mechanical scenario, where the players are represented by different subsystems, and the basis states of 
the subsystems are different choices. To remind, for the KPR game, each of $\lambda N$ customers chooses a restaurant for getting the lunch from $N$ different 
choices in parallel decision mode. The players (customers) receive a payoff if their choice is not too crowded, i.e., the number 
of customers with the same choices is under some threshold limit. For this problem, this limit is considered as one. If more than 
one customer arrives at any restaurant for their lunch, then one of them is randomly chosen to provide the service, and the others 
will not get lunch on that day. 

Let us consider a simple case of three players, say, Alice, Bob and Charlie who have three possible choices: restaurant $1$, 
restaurant $2$ and restaurant $3$. They receive a payoff $\$=1$ if they make a unique choice, otherwise they receive a 
payoff $\$=0$. Therefore, it is a one shoot game, i.e., non-iterative, and the players do not have any knowledge from previous 
rounds to decide their actions. Since the players can not communicate, there is no other way except randomizing between the 
choices. In this case, there are a total $27$ different combinations of choices and $12$ of that provide a payoff $\$=1$ to 
each of the players. Therefore, randomization between the choices leads to an expected payoff $E_c(\$)=4/9$ for each of 
the players using the classical strategy.

The quantum version of the KPR problem with three players ($M=3$; $\lambda=1$) and three choices ($N=3$) is first 
introduced by Sharif and Heydari~\cite{sharif2011quantum}. 
In this case, Alice, Bob and Charlie share a quantum resource. Each of these players has a part in a multipartite quantum 
state. Whereas the classical players are allowed to randomize between their discrete set of choices, for the quantum 
version, each subsystem is allowed to be transformed by local quantum operations. Therefore, choosing a strategy or choice 
is equivalent to choose an unitary operator $U$. In absence of the entanglement in the initial state, it has been found that 
quantum games yield the same payoffs as its classical counterpart. On the other hand, it has been shown that sometimes
a combination of unitary operators and entanglement outperform the classical randomization strategy.  

In this particular KPR problem, the players have three choices, therefore, we need to deal with qutrits instead of qubits 
that are used for two choices to apply quantum protocols~\cite{sharif2012introduction}. The local quantum operations on qutrits are performed by a 
complicated group of matrices from ${\rm SU}(3)$ group, unlike to the case of qubits where the local operators belong 
to ${\rm SU}(2)$ group. A qutrit is a three-level quantum system on three-dimensional Hilbert space 
$\mathcal{H_\mathcal{L}}=\textbf{C}^{3}$. The most general form of quantum state of a qutrit in the computational basis 
is given by
\begin{equation}
  |\psi\rangle=c_{0}|0\rangle+c_{1}|1\rangle+c_{2}|2\rangle,
\end{equation}
where $c_0$, $c_1$ and $c_2$ are three complex numbers satisfying the relation $|c_{0}|^{2}+|c_{1}|^{2}+|c_{2}|^{2}=1$. 
The basis states follow the orthonormal condition $\langle i|j\rangle=\delta_{i,j}$, where $i,j=0, 1, 2$. Then, the general 
state of a $n$-qutrit system can be written as a linear combination of $3^n$ orthonormal basis vectors:
\begin{equation}
  \left|\Psi\right\rangle =\sum_{y_{n},..,y_{1}=0}^{2}c_{y_{n}...y_{1}}\left|y_{n}\cdots y_{1}\right\rangle,
\end{equation}
where the basis vectors are the tensor product of individual qutrit states, defined as, 
\begin{equation}
\left|y_{n}\cdots y_{1}\right\rangle =\left|y_{n}\right\rangle \otimes\left|y_{n-1}\right\rangle \otimes\cdots\otimes\left|y_{1}\right\rangle \in\mathcal{H_\mathcal{L}}=\overbrace{\textbf{C}^{3}\otimes...\otimes\textbf{C}^{3}}^\textrm{$n$-times},
\end{equation}
with $y_{i}\in\{0,1,2\}$. The complex coefficients satisfy the normalization condition $\sum|c_{y_{n}...y_{1}}|^{2}=1$.

A single qutrit can be transformed by an unitary operator $U$ that belongs to special unitary group of degree $3$, denoted 
by SU($3$). In a system of $n$ qutrits, when an operation is performed only on a single qutrit, it is said to be local. 
The corresponding operation changes the state of that particular qutrit only. Under local operations, the state vector 
of a muti-qutrit system is transformed by the tensor products of individual operators, and the final state is given by 
\begin{equation}
\left|\Psi_{\rm fin}\right\rangle =U_{n}\otimes U_{n-1}\otimes\cdots\otimes U_{1}\left|\Psi_{\rm in}\right\rangle,
\end{equation} 
where $|\Psi_{\rm in}\rangle$ is the initial state of the system. 

The SU($3$) matrices, i.e., $3\times3$ unitary matrices are parameterized by defining three orthogonal complex unit 
vectors $\bar{u}, \bar{v}, \bar{w}$, such that $\bar{u}\cdot\bar{v}=0$  and $\bar{u}^*\times\bar{v}=\bar{w}$~\cite{mathur2001coherent}.
A general complex vector with unit norm is given by
\begin{equation}
\bar{u}=\left(\begin{array}{c}
\sin\theta\cos\phi e^{i\alpha_{1}}\\
\sin\theta\sin\phi e^{i\alpha_{2}}\\
\cos\theta e^{i\alpha_{3}}
\end{array}\right),
\end{equation}
where $0\leq\phi,\theta\leq\pi/2$ and $0\leq\alpha_1,\alpha_2,\alpha_3\leq2\pi$. An another complex unit vector 
satisfying $\bar{u}\cdot\bar{v}=0$ is given by
\begin{equation}
\bar{v}=\left(\begin{array}{c}
\cos\chi\cos\theta\cos\phi e^{i(\beta_{1}-\alpha_{1})}+\sin\chi\sin\phi e^{i(\beta_{2}-\alpha_{1})}\\
\cos\chi\cos\theta\sin\phi e^{i(\beta_{1}-\alpha_{2})}-\sin\chi\cos\phi e^{i(\beta_{2}-\alpha_{2})}\\
-\cos\chi\sin\theta e^{i(\beta_{1}-\alpha_{3})}
\end{array}\right),
\end{equation}
where $0\leq\chi\leq\pi/2$ and $0\leq\beta_1,\beta_2\leq2\pi$. The third complex unit vector $\bar{w}$ is determined 
from the orthogonality condition of the complex vectors. Then, a general SU($3$) matrix is constructed 
by placing $\bar{u}, \bar{v}^*$ and $\bar{w}$ as its columns~\cite{mathur2001coherent}, and it can be written as
\begin{equation}
U=\left(\begin{array}{ccc}
u_{1} & v_{1}^{*} & u_{2}^{*}v_{3}-v_{3}^{*}u_{2}\\
u_{2} & v_{2}^{*} & u_{3}^{*}v_{1}-v_{1}^{*}u_{3}\\
u_{3} & v_{3}^{*} & u_{1}^{*}v_{2}-v_{2}^{*}u_{1}
\end{array}\right).
\end{equation}
Therefore, this $3\times3$ matrix is defined by eight real parameters $\phi,\theta,\chi,\alpha_1,\alpha_2,\alpha_3,\beta_1,\beta_2$.

To start the game, we need to choose an initial state that is shared by the players. It can be assumed that 
an unbiased referee prepares the initial state and distributes the subsystems among the players. Thenceforth, no 
communication or interaction is allowed between the players and the referee. To choose an initial state, we need to 
fulfill three criteria: (a) The state should be entangled so that it can accommodate correlated randomization between 
the players. (b) The state must be symmetric and unbiased with respect to the positions of the players, since the game 
follows these properties. (c) It must have the property of accessing the classical game through the restrictions on the 
strategy sets. A state that fulfills these criteria is given by 
\begin{equation}\label{eq:GHZ}
\mid\psi_{in}\rangle=\frac{1}{\sqrt{3}}\left(|000\rangle+|111\rangle+|222\rangle\right).
\end{equation}
This is also a maximally entangled GHZ-type state that is defined on
$\mathcal{H_\mathcal{L}}=\mathbb{C}^{3}\otimes\mathbb{C}^{3}\otimes\mathbb{C}^{3}$. We consider this as the initial state 
to start the game. 

To show that the assumed initial state satisfies the above criterion (c), we consider a set of operators 
representing the classical pure strategies that leads to deterministic payoffs when those are applied to the initial 
state $|\psi_{\rm in}\rangle$. This set of operators is given by the cyclic group of order $3$, $C_3$, generated 
by the $3\times3$ matrix
\begin{equation}
 s=\ \left( \begin{array}{ccc}
0 & 0 & 1 \\
1 & 0 & 0 \\
0 & 1 & 0 \end{array} \right),\
 \end{equation}
with the following properties: $s^0=s^3=I$ and $s^2=s^{-1}=s^T$. Then the players choose their classical strategies from 
a set of operators $S=\{s^0,s^1,s^2\}$ with $ s^a\otimes s^b \otimes s^c|000\rangle=|a\, b\, c\rangle$, where $a, b, c\in \{0, 1, 2\}$. 
By acting the set of classical strategies on the initial state $|\psi_{\rm in}\rangle$, we get the final state as

\begin{multline}
s^a\otimes s^b \otimes s^c\frac{1}{\sqrt{3}}\left(|000\rangle+|111\rangle+|222\rangle\right)= \\
= \frac{1}{\sqrt{3}}\left(|0+a\;0+b\;0+c\rangle+|1+a\;1+b\;1+c\rangle+|2+a\;2+b\;2+c\rangle\right).
\end{multline}
It is important to note here, that the superscripts indicate the powers of the generator matrix and the addition 
is modulo $3$. 

To proceed with the quantum game, an initial density matrix is constructed by using the initial state $|\psi_{\rm in}\rangle$ 
and adding a noise term, controlled by the parameter $f$~\cite{schmid2010experimental}. The density matrix can be written as
\begin{equation}
\rho_{in}=f\mid\psi_{in}\rangle\langle\psi_{in}\mid+\frac{1-f}{27}I_{27},
\label{initial_dm}
\end{equation}
where the parameter $f\in[0,1]$ and $I_{27}$ is the $27\times 27$ identity matrix. The parameter $f$ is a measure of the fidelity 
of production of the initial state~\cite{sharif2012introduction,ramzan2012decoherence}. For $f=0$, the initial state is fully random, since 
the corresponding density matrix has zero off-diagonal elements, and non-zero diagonal elements are of equal strength. 
On the other hand, for $f=1$, the initial state is entangled one with zero noise. For the values of $f$ between $0$ and $1$, 
the initial state is entangled with non-zero noise measured by $f$.
Alice, Bob and Charlie will now choose their strategies by considering an unitary operator $U(\phi,\theta,\chi,\alpha_1,\alpha_2,\alpha_3,\beta_1,\beta_2)$, 
and after their actions, the initial state $\rho_{in}$ transforms into the final state
\begin{equation}
\rho_{fin}=U\otimes U\otimes U \rho_{in} U^{\dagger}\otimes U^{\dagger}\otimes U^{\dagger}.
\end{equation}
We assume here the same unitary operator $U$ for all three players, since there is no scope of communications among them. 
Therefore, it is practically impossible to coordinate which operator to be applied by whom.
The next step is to construct a payoff operator $P_i$ for each of the $i$-th player. This is defined as a sum of outer products 
of the basis states for which $i$-th player receives a payoff $\$=1$.
For example, the payoff operator of Alice is given by
 \begin{multline}
 P_{A}=\left(\sum_{y_3,y_2,y_1=0}^{2}|y_{3}y_{2}y_{1}\rangle\langle y_{3}y_{2}y_{1}|,\, y_3\neq y_2, y_3\neq y_1, y_2\neq y_1\right)+\\
 +\left(\sum_{y_3,y_2,y_1=0}^{2}|y_{3}y_{2}y_{1}\rangle\langle y_{3}y_{2}y_{1}|,\, y_3=y_2\neq y_1\right).
 \label{payoff_op}
 \end{multline}
Note that the terms inside the first bracket of the operator represent the scenario when all three players have different 
choices, whereas the second bracket leads to the fact that Alice's choice is different from Bob and Charlie who have same 
choices. In the same way, one can find the payoff matrices for Bob and Charlie. As defined in Eq.~(\ref{expect_payoff}), the 
expected payoff of player $i$ can be calculated as
\begin{equation}
E_i(\$)=\mathrm{Tr\left(\mathit{P}_{i}\rho_{fin}\right)},
\end{equation}
where $i\in \{A, B, C\}$.

\begin{table}\center
\label{Uopttable}
\begin{tabular}{|c|c|c|c|c|c|c|c|c|c|}
\hline
 Parameter  & $\phi$ & $\theta$  & $\chi$ & $\alpha_1$ & $\alpha_2$ & $\alpha_3$ & $\beta_1$ & $\beta_2$\tabularnewline

\hline
 Value  & $\frac{\pi }{4}$ & $\cos ^{-1}\left(\frac{1}{\sqrt{3}}\right)$ & $\frac{\pi }{4}$ & $\frac{5 \pi }{18}$ & $\frac{5 \pi }{18}$ & $\frac{5 \pi }{18}$ & $\frac{\pi }{3}$ & $\frac{11 \pi }{6}$ \tabularnewline
\hline
\end{tabular}
\caption{Paramater values for an optimal unitary operator $U^{\rm opt}$.}
\end{table}

The problem now is to find an optimal strategy, i.e., to determine a general unitary operator 
$U(\phi, \theta, \chi, \alpha_1, \alpha_2, \alpha_3, \beta_1, \beta_2)$ that maximizes the expected payoff. 
In this game, all the players will have same expected payoff for a particular strategy operator, since they do not 
communicate between each other during the process. Therefore the optimization of expected payoff can be done 
with respect to any of the three players. It has been shown in Ref.~\cite{sharif2011quantum}, that there exists an 
optimal unitary operator $U^{\rm opt}$ with the parameter values listed in table $1$, for which one finds a 
maximum expected payoff of $E(\$)=6/9$, assuming a pure initial state ($f=1$; see Eq.~(\ref{initial_dm})). 
Thus, the quantum strategy outperforms classical randomization, and the expected payoff can be increased 
by $50\%$ compared to the classical case where the expected payoff was found to be $E_c(\$)=4/9$.
It also has been shown that $U^{\rm opt}$ is a Nash equilibrium, because no players can increase their payoff 
by changing their individual strategy from $U^{\rm opt}$ to any other strategy $U$ (for details, see Ref.~\cite{sharif2011quantum}).

By applying $U^{\rm opt}$ on the initial state (see Eq.~(\ref{eq:GHZ})), the final state is given by
\begin{multline}
\mid\psi_{fin}\rangle=\frac{1}{3}\left(|000\rangle+|111\rangle+|222\rangle+|012\rangle+|021\rangle+|102\rangle+|120\rangle\right.+\\\left.|201\rangle+|210\rangle\right).
\end{multline}
Note that this is a collection of all the basis states that leads to providing a payoff either to all 
three players or none of them. We see that the optimal strategy profile $U^{\rm opt}\otimes U^{\rm opt}\otimes U^{\rm opt}$ 
becomes unable to get rid of the most undesired basis states $|000\rangle,|111\rangle,|222\rangle$ 
(i.e., no players will receive any payoff) from the final state. This failure is indeed responsible 
for getting expected payoff $E(\$)=6/9$ instead of unity.
For a general noise term $f$ and optimal strategy, the expected payoff can be calculated as 
$E(\$(U^{\rm opt},f))=\frac{2}{9}(f+2)$~\cite{sharif2011quantum}. This general result is compatible with the case 
of $f=1$, and it also reproduces the classical value as $f \rightarrow 0$. 

\subsection{Effect of entanglement}
\label{entanglement}
\begin{figure}
  \includegraphics[scale=0.75]{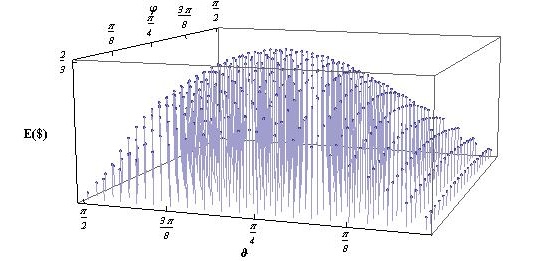}\\
  \caption{Expected payoff $E(\$(U^{\rm opt},\vartheta,\varphi))$ as a function of $\vartheta$ and $\varphi$, for a quantum KPR game with 
  three players and three choices with the optimal strategy operator $U^{\rm opt}$. Each pair of $\vartheta$ and $\varphi$ indicates a different 
  initial state according to Eq.~(\ref{ent_initial_state}). The peak in $E(\$)$ occurs for a maximally entangled initial state, i.e., 
  for $\vartheta=\cos^{-1}(1/\sqrt{3})$ and $\varphi=\pi/4$ (taken from~\cite{sharif2012strategies_arxiv}).}\label{b}
\end{figure}

We now investigate whether the level of entanglement of the initial state affects the payoffs of the players in quantum KPR 
problem with three players and three choices. 
To show this effect, one can start the game with a general entangled state
\begin{equation}
\mid\psi_{in}\rangle=\sin\vartheta\cos\varphi|000\rangle+\sin\vartheta\sin\varphi|111\rangle+\cos\vartheta|222\rangle,
\label{ent_initial_state}
\end{equation}
where $0\leq\vartheta\leq\pi$ and $0\leq\varphi\leq2\pi$. Using the given optimal strategy $U^{\rm opt}$ and the 
above general initial state, the expected payoff can be found as
\begin{equation}
E(\$(U^{opt},\vartheta,\varphi))=\frac{1}{9}\left(\sin(\varphi)\sin(2\vartheta)+\cos(\varphi)\left(2\sin(\varphi)\sin^{2}(\vartheta)+\sin(2\vartheta)\right)+4\right).
\end{equation}
This relation is used to find the values of $\vartheta$ and $\varphi$ for which the expected payoff becomes 
maximum. In Refs.~\cite{sharif2011quantum,sharif2012strategies}, it has been shown that the maximum expected payoff occurs for 
$\varphi=\frac{\pi}{4},\frac{3\pi}{4}$ and $\vartheta=\pm\cos^{-1}(1/\sqrt{3})$, i.e., when the initial state 
is maximally entangled that we have considered in Eq.~(\ref{eq:GHZ}). A small deviation from maximal 
entangled state reduces the expected payoff from its maximum value (see Fig.~\ref{b}). It can be noted here that 
the expected payoff has a strong dependence on the level of entanglement of the initial state.

\subsection{Effect of decoherence}
\label{decoherence}
It is practically impossible to completely isolate a quantum system from the effects of the environment. Therefore, the studies 
that account for such effects have practical implications. In this context, the study of decoherence (or loss of phase information) 
is essential to understand the dynamics of a system in presence of system-environment interactions. Quantum games are recently being explored 
to implement quantum information processing in physical systems~\cite{pakula2007quantum} and can be used to study the effect of decoherence in 
such systems~\cite{flitney2004quantum,johnson2001playing,ramzan2008noise,chen2003quantum,ramzan2010distinguishing}. 
In this connection, different damping channels can be used as a theoretical framework to study the influence of decoherence in quantum game problems.

We here study the effect of decoherence in three-player and three-choice quantum KPR problem by assuming different noise models, such as, 
amplitude damping, phase damping, depolarizing, phase flip and trit-phase flip channels, parameterized by a decoherence parameter $p$,
where $p\in \lbrack 0,1]$~\cite{ramzan2013three}. The lower limit of decoherence parameter represents a completely coherent system, whereas the upper limit 
represents the zero coherence or fully decohered case. 

In a noisy environment, the Kraus operator representation can be used to describe the evolution of a quantum state 
by considering the super-operator $\Phi$~\cite{nielsen2001quantum}. Using density matrix representation, the evolution 
of the state is given by

\begin{equation}
\tilde{\rho}_{f}=\Phi (\rho _{f})=\sum_{k}E_{k}\rho _{f}E_{k}^{\dag },
\label{E55}
\end{equation}%
where the Kraus operators $E_{k}$ follow the completeness relation,
\begin{equation}
\sum_{k}E_{k}^{\dag }E_{k}=I.  \label{5}
\end{equation}%
The Kraus operators for the game are constructed using the single qutrit
Kraus operators as provided in Eqs.~(\ref{E1},\ref{E2},\ref{E3},\ref{E4},\ref{E5}) by taking their tensor
product over all $n^{3}$ combination of $\pi \left( i\right) $ indices

\begin{equation}
E_{k}=\underset{\pi }{\otimes }e_{\pi \left( i\right) },  \label{6}
\end{equation}%
with $n$ being the number of Kraus operators for a single qutrit channel. 
For the amplitude damping channel, the single qutrit Kraus operators are given by~\cite{ann2009finite}

\begin{equation}
E_{0}=\left(
\begin{array}{ccc}
1 & 0 & 0 \\
0 & \sqrt{1-p} & 0 \\
0 & 0 & \sqrt{1-p}%
\end{array}%
\right) ,\ \ E_{1}=\left(
\begin{array}{ccc}
0 & \sqrt{p} & 0 \\
0 & 0 & 0 \\
0 & 0 & 0%
\end{array}%
\right) ,\ \ E_{2}=\left(
\begin{array}{ccc}
0 & 0 & \sqrt{p} \\
0 & 0 & 0 \\
0 & 0 & 0%
\end{array}%
\right).  \label{E1}
\end{equation}%
In a similar way, the single qutrit Kraus operators for the phase damping channel
can be found as~\cite{ramzan2012decoherence}

\begin{equation}
E_{0}=\sqrt{1-p}\left(
\begin{array}{ccc}
1 & 0 & 0 \\
0 & 1 & 0 \\
0 & 0 & 1%
\end{array}%
\right) ,\ \ E_{1}=\sqrt{p}\left(
\begin{array}{ccc}
1 & 0 & 0 \\
0 & \omega & 0 \\
0 & 0 & \omega ^{2}%
\end{array}%
\right) ,  \label{E2}
\end{equation}%
where $\omega =e^{\frac{2\pi i}{3}}.$ 
For the depolarizing channel, the single qutrit Kraus operators take the forms as~\cite{salimi2009investigation}

\begin{equation}
E_{0}=\sqrt{1-p}I,\ E_{1}=\sqrt{\frac{p}{8}}Y,\ E_{2}=\sqrt{\frac{p}{8}}Z,\
E_{3}=\sqrt{\frac{p}{8}}Y^{2},\ E_{4}=\sqrt{\frac{p}{8}}YZ,
\label{E3}
\end{equation}

\begin{equation}
E_{5}=\sqrt{\frac{p}{8}}Y^{2}Z,\ E_{6}=\sqrt{\frac{p}{8}}YZ^{2},\ \ E_{7}=%
\sqrt{\frac{p}{8}}Y^{2}Z^{2},\ \ E_{8}=\sqrt{\frac{p}{8}}Z^{2},  \label{E4}
\end{equation}%
where

\begin{equation}
Y=\left(
\begin{array}{ccc}
0 & 1 & 0 \\
0 & 0 & 1 \\
1 & 0 & 0%
\end{array}%
\right) ,\ \ Z=\left(
\begin{array}{ccc}
1 & 0 & 0 \\
0 & \omega & 0 \\
0 & 0 & \omega ^{2}%
\end{array}%
\right).  \label{E5}
\end{equation}%
The single qutrit Kraus operators, associated with the phase flip channel are given by

\begin{equation}
E_{0}=\left(
\begin{array}{ccc}
1 & 0 & 0 \\
0 & \sqrt{1-p} & 0 \\
0 & 0 & \sqrt{1-p}%
\end{array}%
\right) ,\ \ E_{1}=\left(
\begin{array}{ccc}
0 & \sqrt{p} & 0 \\
0 & 0 & 0 \\
0 & 0 & 0%
\end{array}%
\right) ,\ \ E_{2}=\left(
\begin{array}{ccc}
0 & 0 & \sqrt{p} \\
0 & 0 & 0 \\
0 & 0 & 0%
\end{array}%
\right).
\end{equation}%
Similarly, the single qutrit Kraus operators for the trit-phase flip channel can be found as

\begin{eqnarray}
E_{0} &=&\sqrt{1-\frac{2p}{3}}\left(
\begin{array}{ccc}
1 & 0 & 0 \\
0 & 1 & 0 \\
0 & 0 & 1%
\end{array}%
\right) ,\ \ E_{1}=\sqrt{\frac{p}{3}}\left(
\begin{array}{ccc}
0 & 0 & e^{\frac{2\pi i}{3}} \\
1 & 0 & 0 \\
0 & e^{\frac{-2\pi i}{3}} & 0%
\end{array}%
\right) ,  \notag \\
E_{2} &=&\sqrt{\frac{p}{3}}\left(
\begin{array}{ccc}
0 & e^{\frac{-2\pi i}{3}} & 0 \\
0 & 0 & e^{\frac{2\pi i}{3}} \\
1 & 0 & 0%
\end{array}%
\right) ,\ \ E_{3}=\sqrt{\frac{p}{3}}\left(
\begin{array}{ccc}
0 & e^{\frac{2\pi i}{3}} & 0 \\
0 & 0 & e^{\frac{-2\pi i}{3}} \\
1 & 0 & 0%
\end{array}%
\right),
\end{eqnarray}%
where the term $p=1-e^{-\Gamma t}$ determines the strength of quantum noise which is usually
called as decoherence parameter. This relation describes the bounds $[0,1]$ of $p$ 
by two extreme time limits $t=0$, $\infty $ respectively. 
The final density matrix representing the state after the action of the channel is given by
\begin{equation}
\tilde{\rho _{f}}=\Phi _{p}(\rho _{f})
\end{equation}%
where $\Phi _{p}$ is the super-operator for realizing a quantum channel
parametrized by the decoherence parameter $p$. The payoff
operator for $i^{\text{th}}$ player (say Alice) is given by Eq.~(\ref{payoff_op}).The expected payoff of $i^{\text{th}}$ player can be calculated as
\begin{equation}
E_{i}(\$)=\text{Tr}\{P_{A}\tilde{\rho}_{f}\}
\end{equation}%
where Tr represents the trace of the matrix. We have already studied the zero noise case ($p=0$) in Sec.~\ref{quantum_KPR} 
considering the fidelity $f=1$. It has been found that there exists an optimal unitary operator $U_{\rm opt}$ for which 
the expected payoff of a player becomes maximum. We here consider how a non-zero noise term $p$ and the fidelity,
$f\neq1$ affects the expected payoff.

\begin{figure}[tbp]
\begin{center}
\vspace{-2cm} \includegraphics[scale=0.5]{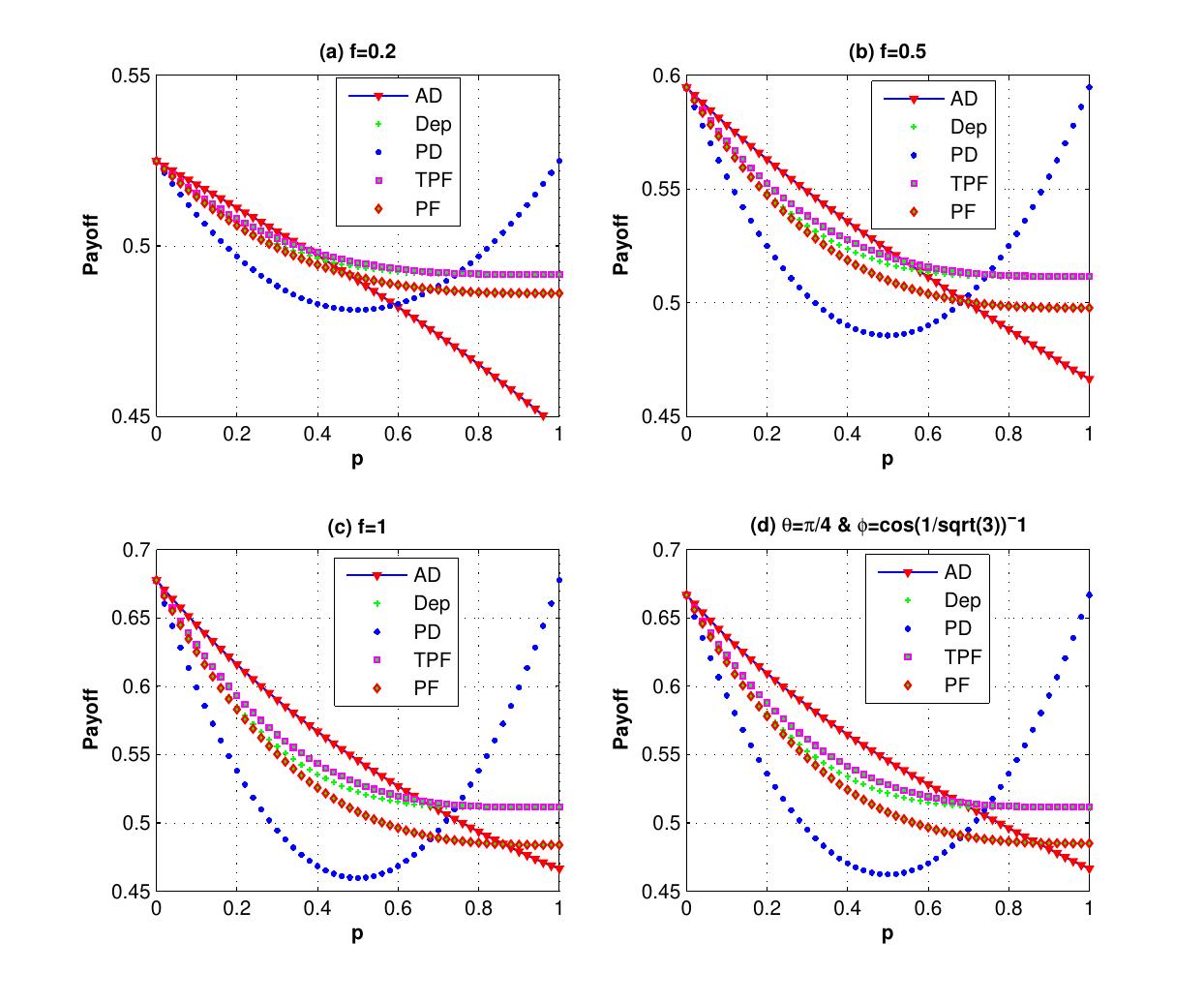} \\[0pt]
\end{center}
\caption{(Color online) Variation of Alice's expected payoff as a function of the
decoherence parameter $p$ for (a) $f=0.2,$ (b) $f=0.5,$ (c) $f=1$ and $%
\protect\theta =\frac{\protect\pi }{4},$ $\protect\phi =\cos ^{-1}(1/\protect%
\sqrt{3})$ for amplitude damping, depolarizing, phase damping, trit-phase
flip and phase flip channels (taken from~\cite{ramzan2013three_arxiv}).}
\label{fig1_deco}
\end{figure}

\begin{figure}[tbp]
\begin{center}
\vspace{-2cm} \includegraphics[scale=0.5]{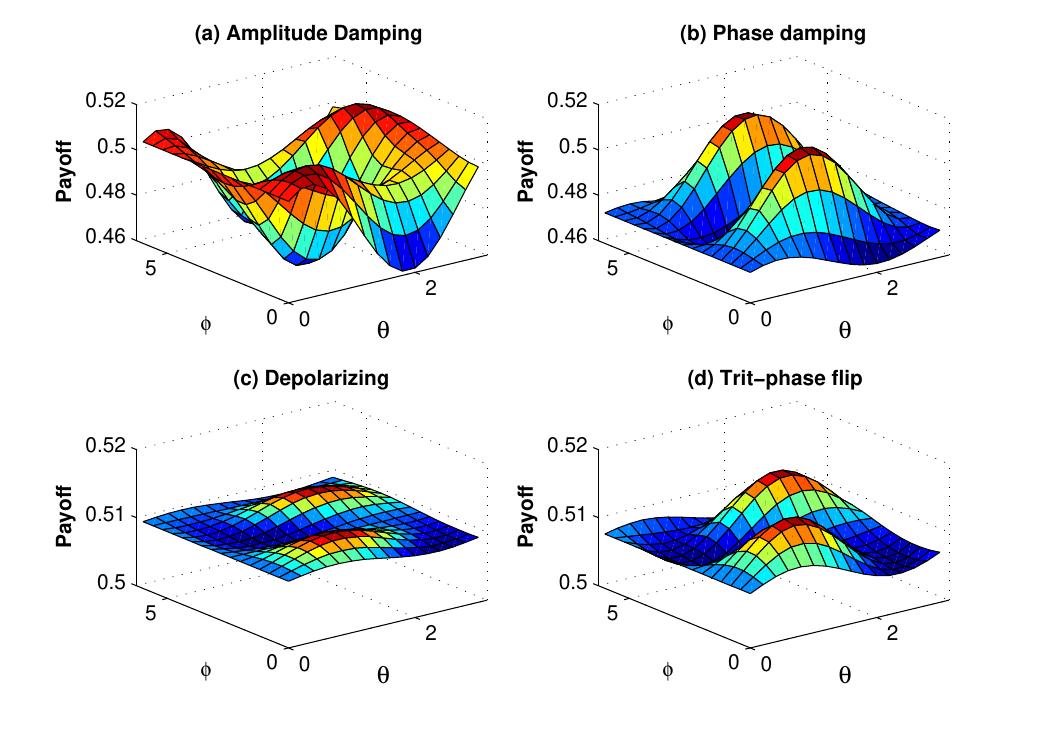} \\[0pt]
\end{center}
\caption{(Color online) Alice's expected payoff as a function of $\protect%
\theta $ and $\protect\phi $ (determined by Eq.~(\ref{ent_initial_state})) for 
(a) amplitude damping, (b) phase damping, (c) depolarizing and (d) trit-phase 
flip channels with decoherence parameter $p=0.7$
 (taken from~\cite{ramzan2013three_arxiv}).}
\label{fig2_deco}
\end{figure}

\begin{figure}[tbp]
\begin{center}
\vspace{-2cm} \includegraphics[scale=0.5]{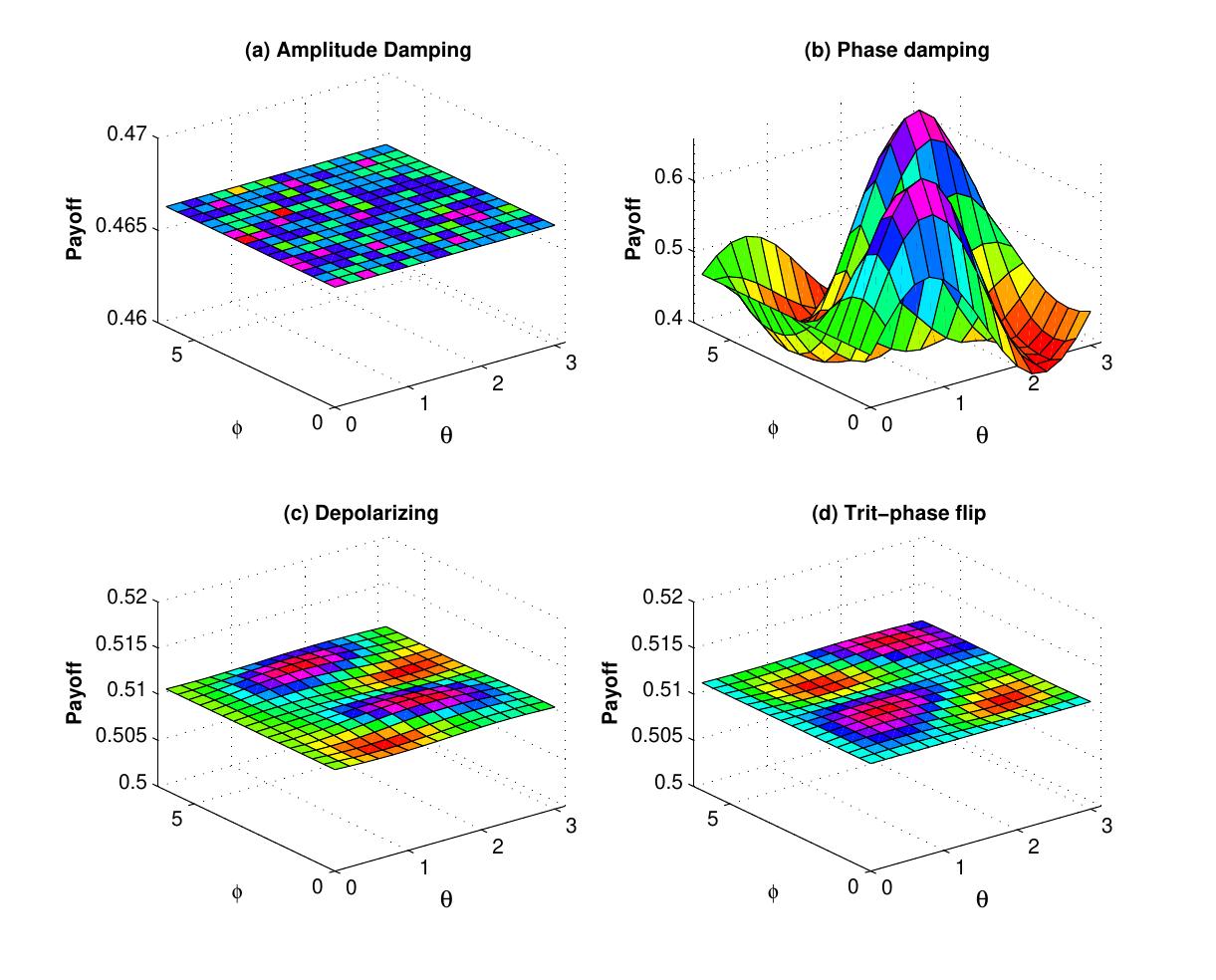} \\[0pt]
\end{center}
\caption{(Color online) Plot of Alice's expected payoff as a function of $\protect%
\theta $ and $\protect\phi $ (determined by Eq.~(\ref{ent_initial_state})) for (a) amplitude damping, (b) phase
damping, (c) depolarizing and (d) trit-phase flip channels with decoherence parameter $p=1$
 (from~\cite{ramzan2013three_arxiv}).}
\label{fig3_deco}
\end{figure}

In order to explain the effect of decoherence on the quantum KPR game, we investigate
expected payoff by varying the decoherence parameter $p$ for different damping channels.
Due to the symmetry of the problem, we have considered expected payoff of one of the
three players (say Alice) for further investigations. In Fig.~\ref{fig1_deco}, the
expected payoff of Alice is plotted as a function of decoherence parameter $p$ for
different values of fidelity $f$ and different damping channels, such as, amplitude
damping, depolarizing, phase damping, trit-phase flip and phase flip channels.
It is observed that Alice's payoff is strongly affected by the amplitude driving
channel as compared to the flipping and depolarizing channels. The effect of
entanglement of initial state is further investigated by plotting Alice's
expected payoff as a function of $\theta$ and $\phi$ in presence of noisy
environment with decoherence parameter $p=0.7$ for different damping cases:
(a) amplitude damping, (b) phase damping, (c) depolarizing and (d) trit-phase
flip channels (see Fig.~\ref{fig2_deco}). In this scenario, one can see that
Alice's payoff is heavily affected by depolarizing noise compared to the other
noise cases. This plot is also repeated for the highest level of decoherence,
i.e., $p=1$ (see Fig.~\ref{fig3_deco}). It is seen that there is a considerable
amount of reduction in Alice's payoff for amplitude damping, deploarizing
and trit-phase flip cases, whereas phase damping channel almost does not
affect the payoff of Alice. Interestingly, the problem becomes noiseless
for the maximum decoherence in the case of phase damping channel. Finally,
the maximum payoff is achieved for the case of highest initial entanglement
and zero noise, and it starts decreasing when the degree of entanglement
deviates from maxima or introduces a non-zero decoherence term.
Moreover, it has also been checked that the introduction of decoherence does
not change the Nash equilibrium of the problem.

\section{Summary and discussions}


In the Kolkata Paise Restaurant or KPR game $\lambda N$ players
choose every day independently but based on past experience
or learning one of the $N~(\rightarrow~\infty)$ restaurants
in the city. As mentioned, the game becomes trivial if a
non-playing dictator prescribes the moves  to each player.
Because of iterative learning, the KPR game is not necessarily
one-shot one, though for random choice (no memory or learning
from the past) by the players.
 
For random choices of restaurants by the players, the game effectively becomes one-shot with
convergence time $\tau = 1$ and  steady utilization  fraction
$f = 1 - exp(-\lambda) \simeq 0.63$~\cite{chakrabarti2009kolkata}, as shown through Eqs.~\ref{eq_KPR_I},~\ref{eq_KPR_II},~\ref{eq_KPR_III} 
 of section~\ref{classical_strategies}. With iterative learning following Eq.~\ref{eq_KPR_V},~\ref{eq_KPR_VI} 
 for $\lambda = 1$, it was shown numerically, as well as
with a crude approximation in~\cite{ghosh2010statistics}, that the utilization
fraction $f$ becomes of order $0.8$ within a couple of iterations
($\tau$ of the order of 10). In~\cite{sinha2020phase} the authors demonstrated
numerically that for $\lambda = 1$,  $f$ can approach unity
when $\alpha$ becomes $0_+$ from above. However the convergence
time $\tau$ at this critical point diverges due to critical
slowing down (see e.g.,~\cite{stanley1972introduction}), rendering such critically slow
leaning of full utilization is hard to employ for practical purposes~\cite{san2020introduction}. The  cases
of  $\lambda < 1$ (and $\alpha = 1$) were considered earlier in~\cite{ghosh2012phase}and \cite{sinha2020phase}, and has also been studied here  in section~\ref{classical_strategies},
using Monte Carlo technique.  A mean field-like transition
(see e.g.,~\cite{stanley1972introduction}) is observed here giving full utilization
($f = 1$) for $\lambda$ less than $\lambda_c$ about 0.75
where $\tau$ also remains finite. As shown in~Fig.~\ref{fig_I},
$\tau$ diverges at the critical point  $\lambda_c$ (a crude
mean field derivation of it is given in Eq.~\ref{eq_KPR_VII}).

For the quantum version of the KPR problem, we have discussed the one-shot game with
three players and three choices that was first introduced in~\cite{sharif2011quantum}; see 
also \cite{sharif2012introduction}. For this particular KPR game with three players and three choices, 
the classical randomization provides a total $27$ possible configurations, and $12$ out of them gives 
a payoff $\$=1$ to each of the players, thus leading to an expected payoff $E_c(\$)=4/9$.
On the other hand, for the quantum case, it has been shown that when the players share a maximally entangled initial state, 
there exists a local unitary operation (same for all the players due to the symmetry of the problem) 
for which the players receive a maximum expected payoff $E_q(\$)=6/9$, i.e., the quantum players can increase their 
expected payoff by $50\%$ compared to their classical counterpart. To show the effect of entanglement, 
the expected payoff is calculated for a general GHZ-type initial state with different levels of entanglement (see Fig.~\ref{b}). 
It appears that the maximally entangled initial state provides maximum payoff $E_q(\$)=6/9$ to each of 
the players. The expected payoff decreases from its maximum value for any deviation from the maximum 
entanglement of the initial state (see Fig.~\ref{b}). This is the highest expected payoff that is attained 
so far for one-shot quantum KPR game with three players and three choices (see~\cite{sharif2012strategies}). 
Until now, the study of the quantum KPR problem is limited to one shot with three players and three choices, 
and no attempt is found yet to make it iterative, and also to increase the number of players and choices. 
Therefore, it is yet to be understood whether one can increase the expected payoff by making the quantum 
KPR game iterative with learning from previous rounds as happened in the case of classical strategies 
studied here.

Decoherence is an unavoidable phenomenon for quantum systems, since it is not possible to completely 
isolate a system from the effects of the environment. Therefore, it is important to investigate 
the influence of decoherence on the payoffs of the players in the context of quantum games. The 
effect of decoherence in a three-player and three-choice quantum KPR problem is studied 
in \cite{ramzan2013three} using different noise models like amplitude damping, phase damping, depolarizing, 
phase flip and trit-phase flip channels, parametrized by a decoherence parameter. The lower and upper limits 
of decoherence parameter represent the fully coherent and fully decohered system, respectively.
Expected payoff is reported to be strongly affected by amplitude damping channel as compared to the flipping and depolarizing channels
for lower level of decoherence, whereas it is heavily influenced by depolarizing noise in case of higher level of decoherence.
However, for the case of highest level of decoherence, amplitude damping channel dominates over 
the depolarizing and flipping channels, and the phase damping channel has nearly no effect on the payoff.
Importantly, the Nash equilibrium of the problem is shown not to be changed under
decoherence.


There have been several applications of KPR game strategies to
various social optimization cases. KPR game has been extended
to Vehicle for Hire Problem  in~\cite{martin2017extending,martin2017vehicle}. Authors have
built several model variants such as Individual Preferences,
Mixed Preferences, Individual Preferences with Multiple
Customers per District, Mixed Preferences and Multiple
Customers per District, etc.. Using these variants, authors
have studied various strategies for the KPR problem that
led to a foundation of incentive scheme for dynamic matching
in mobility markets. Also in~\cite{martin2017vehicle}, a modest level of
randomization in choice along with mixed strategies is shown
to achieve around 80\% of efficiency in vehicle for hire
markets. A time-varying location specific resource allocation
crowdsourced transportation is studied using the
methodology of mean field equilibrium~\cite{yang2018mean}. This study
provides a detailed mean field analysis of the  KPR game and
also considers the implications of an additional reward
function.  In~\cite{park2017kolkata}, a resource allocation problem for
large Internet-of-Things (IoT) system, consisting of IoT
devices with imperfect knowledge,  is formulated using
the KPR game strategies. The solution, where those IoT
devices autonomously learn  equilibrium strategies to
optimize their transmission, is shown to coincides with
the Nash equilibrium. Also, several `emergent properties'
of the KPR game, such as the utilization fraction or the
occupation density in the steady  or stable states, phase
transition behavior, etc.,  have been numerically studied
in~\cite{tamir2018econophysics}. A search problem often arises as a
time-limited business opportunity by business firms, and
is studied like an one-shot delegated search in~\cite{hassin2021equilibrium}. Authors
here had discussed and investigated the searchers’ incentives
following different strategies, including KPR, that can
maximize the search success. Authors in~\cite{kastampolidou2021dkprg} have discussed
KPR problem as Traveling Salesman Problem (TSP) assuming
restaurants are uniformly distributed on a two-dimensional
plane and this topological layout of the restaurants can
help provide each agent a second chance for choosing a better
or less crowded restaurant. And they have proposed a
meta-heuristics, producing near-optimal solutions in finite
time (as exact solutions of the TSP are prohibitively expensive).
Thus agents are shown to learn fast, even incorporating their
own preferences, and achieves maximum social utilization in
lesser time with multiple chances.

\section*{Acknowledgement}
We are grateful to  our colleagues Anindya
Chakrabarti, Arnab Chatterjee, Daniele  De Martino,
Asim Ghosh, Matteo Marsili, Manipushpak  Mitra and Boaz Tamir
for their collaborations at various stages of the
development of this study. BKC is also grateful to
Indian National Science Academy for their Research
Grant. AR acknowledges UGC, India for start-up research grant F.
30-509/2020(BSR).



\end{document}